\documentclass[12pt]{extarticle}
\usepackage[english]{babel}
\usepackage[margin=2.54cm, letterpaper]{geometry}
\usepackage[english]{babel}
\usepackage[utf8x]{inputenc}
\usepackage[T1]{fontenc} 
\usepackage{amsmath}
\usepackage{bm}
\usepackage{graphicx}
\usepackage[colorinlistoftodos,prependcaption,textsize=small]{todonotes}
\usepackage{xargs}                      % Use more than one optional parameter in a new commands
\usepackage{filecontents}
\usepackage{natbib}		
\usepackage[pdfencoding=unicode]{hyperref}
\usepackage{parskip} 
\usepackage{colortbl}
\usepackage{tabularx}
\usepackage{titlesec}
\usepackage[titletoc, title]{appendix}
\usepackage{subfigure}
\usepackage{boldline, multirow}
\usepackage{array}
\usepackage{arydshln}
\usepackage{cprotect}
\usepackage{adjustbox}
\usepackage{pdflscape}
\usepackage{times}
\usepackage{titlesec}
\usepackage{fancyhdr}
\usepackage[hang,flushmargin,perpage]{footmisc}
\usepackage[normalem]{ulem}
\usepackage{nicefrac}
\usepackage{amsbsy}
\usepackage{afterpage}
\usepackage{enumitem}		% This I used to adjust the left anf right margins when I use enumerate. It works with [leftmargin=2cm]
\usepackage{datetime}		% This is to customise the date in the title page.
\usepackage{amssymb} 		% This is to use the R operator for the real set of numbers.
\usepackage{makecell} 		% This is to use \thead and input two lines in one cell in a table
\usepackage{hhline} 			% This is to create a double line in a table in the form of =======
\usepackage{lineno}			% This is combined with the \linenumbers command
\usepackage{relsize}			% This is to increase the size of the integral in mathmode
\usepackage[skip=0.5\baselineskip]{caption}		% This is to define the vertical space between the caption and the table or the figure
\newdateformat{monthyeardate}{%				% This is to customise the date in the title page.
  \monthname[\THEMONTH] \THEYEAR}

\usepackage[nodisplayskipstretch]{setspace}		% Sets variable line spacings. This package provides the commands \doublespacing, \onehalfspacing, \singlespacing and \setstretch{baselinestretch}, which will specify the line spacing for all sections and paragraphs until another command is used.
\hypersetup{
    colorlinks=true,
    linkcolor=blue,
    citecolor=blue,
    filecolor=magenta,
    urlcolor=blue,
}

\bibpunct{(}{)}{,}{a}{}{,~}

\titleformat{\section}
  {\normalfont\fontsize{14}{14}\bfseries}{\thesection}{1em}{}

\titleformat{\paragraph}
{\normalfont\fontsize{12}{12}\bfseries}{\theparagraph}{1em}{}
\titlespacing*{\paragraph}
{0pt}{3.25ex plus 1ex minus .2ex}{1.5ex plus .2ex}

\newcommandx{\unsure}[2][1=]{\todo[linecolor=red,backgroundcolor=red!25,bordercolor=red,#1]{#2}}
\newcommandx{\change}[2][1=]{\todo[linecolor=blue,backgroundcolor=blue!25,bordercolor=blue,#1]{#2}}
\newcommandx{\info}[2][1=]{\todo[linecolor=green,backgroundcolor=green!25,bordercolor=green,#1]{#2}}
\newcommandx{\improvement}[2][1=]{\todo[linecolor=pink,backgroundcolor=pink!25,bordercolor=pink,#1]{#2}}
\newcommandx{\thiswillnotshow}[2][1=]{\todo[disable,#1]{#2}}
\newcommandx{\checkthat}[2][1=]{\todo[linecolor=gray,backgroundcolor=gray!25,bordercolor=gray,#1]{#2}}

\newcommand{\white}[1]{\textcolor{white}{#1}}

\presetkeys{todonotes}{size=\setstretch{0.8}\selectfont}{}
\reversemarginpar


\newcolumntype{L}[1]{>{\raggedright\arraybackslash}p{#1}}  % This defines a table with fixed column widths
\newcolumntype{C}[1]{>{\centering\arraybackslash}p{#1}} % This defines a table with fixed column widths
\newcolumntype{R}[1]{>{\raggedleft\arraybackslash}p{#1}} % This defines a table with fixed column widths
\newcolumntype{Y}{>{\centering\arraybackslash}X}

\usepackage{accents}

\newcommand{\0}{\text{0}}
\newcommand{\1}{\text{1}}
\newcommand{\2}{\text{2}}
\newcommand{\3}{\text{3}}
\newcommand{\comma}{\text{,}}

\title{ \Large \textbf{{\singlespacing The fractional derivative of the Dirac delta function and new results on the inverse Laplace transform of irrational functions}}}
\author{\vspace{-5ex}}
\date{\vspace{-5ex}}

\makeatletter
\renewenvironment{abstract}{%
    \if@twocolumn
      \section*{\abstractname}%
    \else %% <- here I've removed \small
      \begin{center}%
        {\bfseries \Large\abstractname\vspace{-0.5cm}}%  %% <- here I've added \Large
      \end{center}%
      \quotation
    \fi}
    {\if@twocolumn\else\endquotation\fi}
\makeatother

\begin{document}

%\linenumbers
\maketitle
\vspace*{-1.5cm}

\begin{center}
\singlespacing
{\large Nicos Makris}\footnote{
\textit{Department of Civil and Environmental Engineering, Southern Methodist University, Dallas, Texas, 75276} \\
\vspace*{0.25cm} 
nmakris@smu.edu \\
\textit{Office of Theoretical and Applied Mechanics, Academy of Athens, 10679, Greece}}
\onehalfspacing
\end{center}

\vspace*{0.5cm}
\begin{abstract}
\singlespacing
{\small %
\noindent Motivated from studies on anomalous diffusion, we show that the memory function $M(t)$ of complex materials, that their creep compliance follows a power law, $J(t)\sim t^q$ with $q\in \mathbb{R}^+$, is the fractional derivative of the Dirac delta function, {\normalsize $\frac{\mathrm{d}^q\delta(t-\0)}{\mathrm{d}t^q}$} with $q\in \mathbb{R}^+$. This leads to the finding that the inverse Laplace transform of $s^q$ for any $q\in \mathbb{R}^+$ is the fractional derivative of the Dirac delta function, {\normalsize $\frac{\mathrm{d}^q\delta(t-\0)}{\mathrm{d}t^q}$}. This result, in association with the convolution theorem, makes possible the calculation of the inverse Laplace transform of {\normalsize $\frac{s^q}{s^{\alpha}\mp\lambda}$} where $\alpha<q\in\mathbb{R}^+$ which is the fractional derivative of order $q$ of the Rabotnov function {\normalsize $\varepsilon$}$_{\alpha-\1}(\pm\lambda\comma\,t)=t^{\alpha-\1}E_{\alpha\comma\,\alpha}(\pm\lambda t^{\alpha})$. The fractional derivative of order $q\in \mathbb{R}^+$ of the Rabotnov function, {\normalsize $\varepsilon$}$_{\alpha-\1}(\pm\lambda\comma\,t)$ produces singularities which are extracted with a finite number of fractional derivatives of the Dirac delta function depending on the strength of $q$ in association with the recurrence formula of the two-parameter Mittag--Leffler function.}
\onehalfspacing
\end{abstract}

\singlespacing
{\indent \small \textbf{\textsl{Keywords:}} Generalized functions, Laplace transform, anomalous diffusion, fractional calculus, \\ \indent Mittag--Leffler function}
\onehalfspacing

%\vspace*{0.cm}
\section{Introduction}
\vspace*{-0.5cm}
The classical result for the inverse Laplace transform of the function $\mathcal{F}(s)=$ {\large $\frac{\1}{s^{q}}$} is \citep{Erdelyi1954}
\begin{equation}\label{eq:Eq01}
\mathcal{L}^{-\1}\left\lbrace \frac{\1}{s^{q}} \right\rbrace = \frac{\1}{\Gamma(q)}t^{q-\1} \enskip \text{with} \enskip q>\0
\end{equation}
In Eq. \eqref{eq:Eq01} the condition $q>\0$ is needed because when $q=\0$, the ratio {\large $\frac{\1}{\Gamma(\0)}$} $=\0$ and the right-hand side of Eq. \eqref{eq:Eq01} vanishes, except when $t=\0$ which leads to a singularity. Nevertheless, within the context of generalized functions, when $q=\0$, the right-hand side of Eq. \eqref{eq:Eq01} becomes the Dirac delta function \citep{Lighthill1958} according to the \citet{GelfandShilov1964} definition of the $n^{\text{th}}$ $(n \in \mathbb{N}_{\0})$ derivative of the Dirac delta function
\begin{equation}\label{eq:Eq02}
\frac{\mathrm{d}^n\delta(t-\0)}{\mathrm{d}t^n}=\frac{\1}{\Gamma(-n)} \, \frac{\1}{t^{n+\1}} \enskip \text{with} \enskip n \in \left\lbrace \0\comma\,\1\comma\,\2\,  ...   \right\rbrace
\end{equation}
with a proper interpretation of the quotient {\large $\frac{\1}{t^{n+\1}}$} as a limit at $t=\0$. So according to the \citet{GelfandShilov1964} definition expressed by Eq. \eqref{eq:Eq02}, Eq. \eqref{eq:Eq01} can be extended for values of $q \in \left\lbrace \0\comma\, -\1\comma\,-\2\comma\,-\3\,  ...   \right\rbrace$ and in this way one can establish the following expression for the inverse Laplace transform of $s^n$ with $n \in \mathbb{N}_{\0}$
\begin{equation}\label{eq:Eq03}
\mathcal{L}^{-\1}\left\lbrace s^n \right\rbrace = \frac{\1}{\Gamma(-n)} \, \frac{\1}{t^{n+\1}} = \frac{\mathrm{d}^n\delta(t-\0)}{\mathrm{d}t^n} \enskip n \in \left\lbrace \0\comma\,\1\comma\,\2\,  ...   \right\rbrace
\end{equation}
For instance when $n=1$, Eq. \eqref{eq:Eq03} yields
\begin{equation}\label{eq:Eq04}
\mathcal{L}^{-\1}\left\lbrace s \right\rbrace = \frac{\1}{\Gamma(-\1)} \, \frac{\1}{t^{\2}} = \frac{\mathrm{d}\delta(t-\0)}{\mathrm{d}t} 
\end{equation}
which is the correct result, since the Laplace transform of {\large $\frac{\mathrm{d}\delta(t-\0)}{\mathrm{d}t}$} is 
\begin{equation}\label{eq:Eq05}
\mathcal{L}\left\lbrace \frac{\mathrm{d}\delta(t-\0)}{\mathrm{d}t} \right\rbrace = \int_{\0^-}^{\infty} \frac{\mathrm{d}\delta(t-\0)}{\mathrm{d}t} e^{-st}\mathrm{d}t=\left. -\frac{\mathrm{d}(e^{-st})}{\mathrm{d}t} \right\vert_{t=\0} = -(-s) = s
\end{equation}
Equation \eqref{eq:Eq05} is derived by making use of the property of the Dirac delta function and its higher-order derivatives 
\begin{equation}\label{eq:Eq06}
\int_{\0^-}^{\infty}\frac{\mathrm{d}^n\delta(t-\0)}{\mathrm{d}t^n}f(t) = (-\1)^n \frac{\mathrm{d}^nf(\0)}{\mathrm{d}t^n} \enskip \text{with} \enskip n \in \left\lbrace \0\comma\,\1\comma\,\2\,  ...   \right\rbrace
\end{equation}
In Eqs. \eqref{eq:Eq05} and \eqref{eq:Eq06}, the lower limit of integration, $\0^-$ is a shorthand notation for $\lim\limits_{\varepsilon \to \0^+}\displaystyle\int_{-\varepsilon}^{\infty}$ and it emphasizes that the entire singular function {\large $\frac{\mathrm{d}^n\delta(t-\0)}{\mathrm{d}t^n}$} $(n \in \mathbb{N}_{\0})$ is captured by the integral operator. In this paper we first show that Eq. \eqref{eq:Eq03} can be further extended for the case where the Laplace variable is raised to any positive real power; $s^q$ with $q \in \mathbb{R}^+$. This generalization, in association with the convolution theorem allows for the derivation of some new results on the inverse Laplace transform of irrational functions that appear in problems with fractional relaxation and fractional diffusion \citep{Nutting1921, Gemant1936, Gemant1938, Koeller1984, Friedrich1991, SchiesselMetzlerBlumenNonnenmacher1995, Lutz2001, Makris2020}.

Most materials are viscoelastic; they both dissipate and store energy in a way that depends on the frequency of loading. Their resistance to an imposed time-dependent shear deformation, $\gamma(t)$, is parametrized by the complex dynamic modulus $\mathcal{G}(\omega)=$ {\large $\frac{\tau(\omega)}{\gamma(\omega)}$} where $\tau(\omega)=\displaystyle\int_{-\infty}^{\infty}\tau(t)e^{-\operatorname{i}\omega t}\mathrm{d}t$ and $\gamma(\omega)=\displaystyle\int_{-\infty}^{\infty}\gamma(t)e^{-\operatorname{i}\omega t}\mathrm{d}t$ are the Fourier transforms of the output stress, $\tau(t)$, and the input strain, $\gamma(t)$, histories. The output  stress history, $\tau(t)$, can be computed in the time domain with the convolution integral
\begin{equation}\label{eq:Eq07}
\tau(t)=\int_{\0^-}^{t} M(t-\xi)\gamma(\xi)\mathrm{d}\xi
\end{equation}
where $M(t-\xi)$ is the memory function of the material \citep{BirdArmstrongHassager1987, DissadoHill1989, Giesekus1995} defined as the resulting stress at time $t$ due to an impulsive strain input at time $\xi(\xi<t)$, and is the inverse Fourier transform of the complex dynamic modulus
\begin{equation}\label{eq:Eq08}
M(t)=\frac{\1}{\2\pi}\int_{-\infty}^{\infty}\mathcal{G}(\omega)e^{\operatorname{i}\omega t}\mathrm{d}\omega
\end{equation}

\section{The Fractional Derivative of the Dirac Delta Function}
\vspace*{-0.5cm}
Early studies on the behavior of viscoelastic materials that their time-response functions follow power laws have been presented by \citet{Nutting1921}, who noticed that the stress response of several fluid-like materials to a step strain decays following a power law, $\tau(t) \sim t^{-q}$ with $\0 \leq q \leq \1$. Following \citeauthor{Nutting1921}'s observation and the early work of \citet{Gemant1936, Gemant1938} on fractional differentials, \citet{ScottBlair1944, ScottBlair1947} pioneered the introduction of fractional calculus in viscoelasticity. With analogy to the Hookean spring, in which the stress is proportional to the zero-th derivative of the strain and the Newtonian dashpot, in which the stress is proportional to the first derivative of the strain, \citeauthor{ScottBlair1944} and his co-workers (\citeyear{ScottBlair1944, ScottBlair1947, ScottBlairCaffyn1949}) proposed the springpot element --- that is a mechanical element in-between a spring and a dashpot with constitutive law
\begin{equation}\label{eq:Eq09}
\tau(t)=\mu_q \frac{\mathrm{d}^q\gamma(t)}{\mathrm{d}t^q} \comma \enskip \0 \leq q \leq \1 
\end{equation}
where $q$ is a positive real number, $\0 \leq q \leq \1$, $\mu_q$ is a phenomenological material parameter with units $\left[\text{M}\right]\left[\text{L}\right]^{-\1}\left[\text{T}\right]^{q-\2}$ (say \textit{Pa$\cdot$sec}$^q$) and {\large $\frac{\mathrm{d}^q\gamma(t)}{\mathrm{d}t^q}$} is the fractional derivative of order $q$ of the strain history, $\gamma(t)$.

A definition of the fractional derivative of order $q$ is given through the convolution integral
\begin{equation}\label{eq:Eq10}
I^q\gamma(t)=\frac{\1}{\Gamma(q)}\int_{c}^{t}(t-\xi)^{q-\1}\gamma(\xi)\mathrm{d}\xi
\end{equation}
where $\Gamma(q)$ is the Gamma function. When the lower limit, $c=\0$, the integral given by Eq. \eqref{eq:Eq10} is often referred to as the Riemann--Liouville fractional integral \citep{OldhamSpanier1974, SamkoKilbasMarichev1974, MillerRoss1993, Podlubny1998}. The integral in Eq. \eqref{eq:Eq10} converges only for $q>\0$, or in the case where $q$ is a complex number, the integral converges for $\mathcal{R}(q)>\0$. Nevertheless, by a proper analytic continuation across the line $\mathcal{R}(q)=\0$, and provided that the function $\gamma(t)$ is $n$ times differentiable, it can be shown that the integral given by Eq. \eqref{eq:Eq10} exists for $n-\mathbb{R}(q)>\0$ \citep{Riesz1949}. In this case the fractional derivative of order $q\in \mathbb{R}^+$ exists and is defined as
\begin{equation}\label{eq:Eq11}
\frac{\mathrm{d}^{q}\gamma(t)}{\mathrm{d}t^{q}}=I^{-q}\gamma(t)=\frac{\1}{\Gamma(-q)}\int_{\0^-}^{t}\frac{\gamma(\xi)}{(t-\xi)^{q+\1}} \mathrm{d}\xi  \comma \enskip q \in \mathbb{R}^+
\end{equation}
where $\mathbb{R}^+$ is the set of positive real numbers and the lower limit of integration, $\0^-$, may capture an entire singular function at the time origin such as $\gamma(t)=\delta(t-\0)$ \citep{Lighthill1958}. Equation \eqref{eq:Eq11} indicates that the fractional derivative of order $q$ of $\gamma(t)$ is essentially the convolution of $\gamma(t)$ with the kernel {\large $\frac{t^{-q-\1}}{\Gamma(-q)}$} \citep{OldhamSpanier1974, SamkoKilbasMarichev1974, MillerRoss1993, Mainardi2010}. The Riemann--Liouville definition of the fractional derivative of order $q \in \mathbb{R}^+$ given by Eq. \eqref{eq:Eq11}, where the lower limit of integration is zero, is relevant to rheology since the strain and stress histories, $\gamma(t)$ and $\tau(t)$, are causal functions, being zero at negative times.  

The Fourier transform of the fractional derivative of a function defined by Eq. \eqref{eq:Eq11} is
\begin{equation}\label{eq:Eq12}
\mathcal{F}\left\lbrace \frac{\mathrm{d}^q\gamma(t)}{\mathrm{d}t^q} \right\rbrace = \int_{-\infty}^{\infty} \frac{\mathrm{d}^q\gamma(t)}{\mathrm{d}t^q} e^{-\operatorname{i}\omega t}\mathrm{d}t = \int_{\0}^{\infty} \frac{\mathrm{d}^q\gamma(t)}{\mathrm{d}t^q} e^{-\operatorname{i}\omega t}\mathrm{d}t = (\operatorname{i}\omega)^q\gamma(\omega)
\end{equation}
where $\mathcal{F}$ indicates the Fourier transform operator \citep{Erdelyi1954, MillerRoss1993, Mainardi2010}. The one-sided integral appearing in Eq. \eqref{eq:Eq12} that results from the causality of the strain history, $\gamma(t)$ is also the Laplace transform of the fractional derivative of the strain history, $\gamma(t)$
\begin{equation}\label{eq:Eq13}
\mathcal{L}\left\lbrace \frac{\mathrm{d}^q\gamma(t)}{\mathrm{d}t^q} \right\rbrace = \int_{\0}^{\infty} \frac{\mathrm{d}^q\gamma(t)}{\mathrm{d}t^q} e^{-st}\mathrm{d}t = s^q\gamma(s)  
\end{equation}
where $s=\operatorname{i}\omega$ is the Laplace variable and $\mathcal{L}$ indicates the Laplace transform operator \citep{LePage1961, Mainardi2010}.

For the elastic Hookean spring with elastic modulus, $G$, its memory function as defined by Eq. \eqref{eq:Eq08} is $M(t)=G\delta(t-\0)$ -- that is the zero-order derivative of the Dirac delta function; whereas, for the Newtonian dashpot with viscosity, $\eta$,  its memory function is $M(t)=\eta \,${\large$\frac{\mathrm{d}\delta(t-\0)}{\mathrm{d}t}$} -- that is the first-order derivative of the Dirac delta function \citep{BirdArmstrongHassager1987}. Since the springpot element defined by Eq. \eqref{eq:Eq09} with $\0 \leq q \leq \1$ is a constitutive model that is in-between the Hookean spring and the Newtonian dashpot, physical continuity suggests that the memory function of the springpot model given by Eq. \eqref{eq:Eq09} shall be of the form of $M(t)=\mu_q$ {\large$\frac{\mathrm{d}^q\delta(t-\0)}{\mathrm{d}t^q}$} -- that is the fractional derivative of order $q$ of the Dirac delta function \citep{OldhamSpanier1974, Podlubny1998}.

\begin{figure}[b!]
\centering
\includegraphics[width=0.8\linewidth, angle=0]{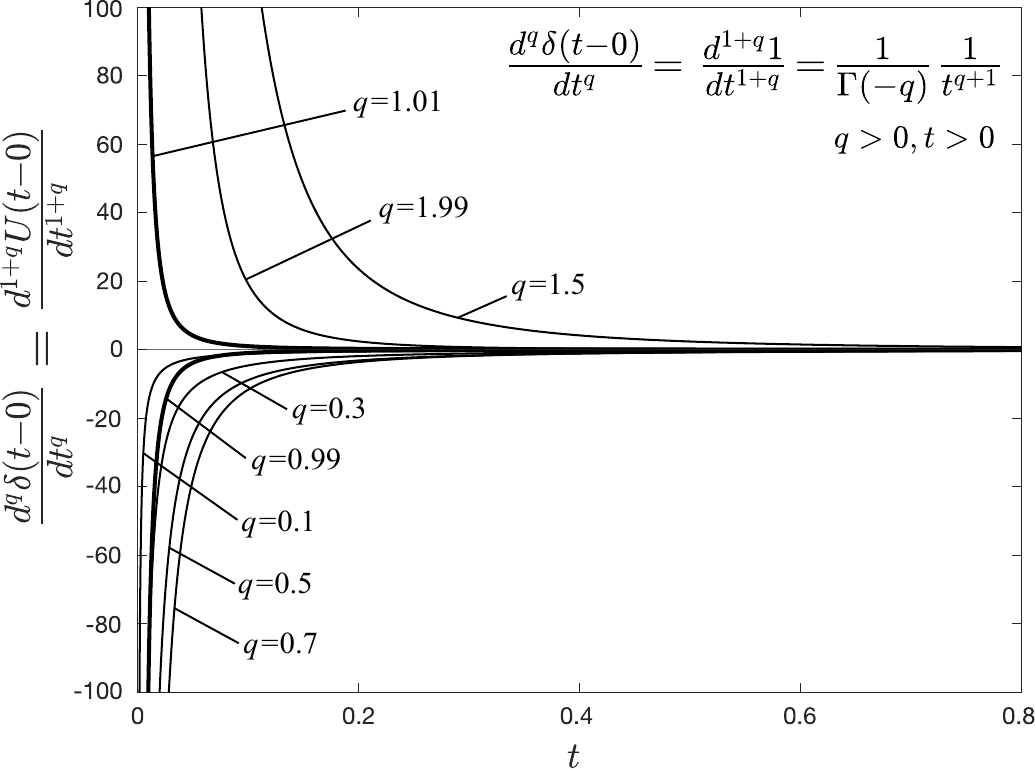}
\caption{Plots of the fractional derivative of the Dirac delta function of order $q\in\left\lbrace\mathbb{R}^+-\mathbb{N}\right\rbrace$, which are the $\1+q$ order derivative of the constant 1 for positive times. The functions are finite everywhere other than the time origin, $t=\0$. Figure \ref{fig:Fig01} shows that the fractional derivatives of the singular Dirac delta function and these of the constant unit at positive times are expressed with the same family of functions.}
\label{fig:Fig01}
\end{figure}

The fractional derivative of the Dirac delta function emerges directly from the property of the Dirac delta function \citep{Lighthill1958}
\begin{equation}\label{eq:Eq14}
\int_{-\infty}^{\infty}\delta(t-\xi)f(t)\mathrm{d}t=f(\xi)
\end{equation}
By following the Riemann--Liouville definition of the fractional derivative of a function given by the convolution appearing in Eq. \eqref{eq:Eq11}, the fractional derivative of order $q\in \mathbb{R}^+$ of the Dirac delta function is
\begin{equation}\label{eq:Eq15}
\frac{\mathrm{d}^q\delta(t-\xi)}{\mathrm{d}t^q}=\frac{\1}{\Gamma(-q)}\int_{\0^-}^t \frac{\delta(\tau-\xi)}{(t-\tau)^{\1+q}}\mathrm{d}\tau \comma \enskip q\in \mathbb{R}^+
\end{equation} 
and by applying the property of the Dirac delta function given by Eq. \eqref{eq:Eq14}; Eq. \eqref{eq:Eq15} gives
\begin{equation}\label{eq:Eq16}
\frac{\mathrm{d}^q\delta(t-\xi)}{\mathrm{d}t^q}=\frac{\1}{\Gamma(-q)}\frac{\1}{(t-\xi)^{\1+q}}  \comma \enskip q\in \mathbb{R}^+
\end{equation}
Equation \eqref{eq:Eq16} offers the remarkable result that the fractional derivative of the Dirac delta function of any order $q\in\left\lbrace\mathbb{R}^+-\mathbb{N}\right\rbrace$ is finite everywhere other than at $t=\xi$; whereas, the Dirac delta function and its integer-order derivatives are infinite-valued, singular functions that are understood as a monopole, dipole and so on; and we can only interpret them through their mathematical properties as the one given by Eqs. \eqref{eq:Eq06} and \eqref{eq:Eq14}. Figure \ref{fig:Fig01} plots the fractional derivative of the Dirac delta function at $\xi=\0$
\begin{equation}\label{eq:Eq17}
\frac{\mathrm{d}^q\delta(t-\0)}{\mathrm{d}t^q}=\frac{\1}{\Gamma(-q)}\frac{\1}{t^{\1+q}} \enskip \enskip \text{with} \enskip  q\in \mathbb{R}^+ \comma \enskip t>\0
\end{equation}
The result of Eq. \eqref{eq:Eq17} for $q \in \mathbb{R}^+$ is identical to the \citet{GelfandShilov1964} definition of the $n^{\text{th}}$ $(n \in \mathbb{N}_{\0})$ derivative of the Dirac delta function given by Eq. \eqref{eq:Eq02} where $\mathbb{N}_{\0}$ is the set of positive integers including zero.

The result for the fractional derivative of the Dirac delta function given by Eq. \eqref{eq:Eq17} is also compared with the well known results in the literature for the fractional derivative of the constant unit function $f(t)=\1$ \citep{OldhamSpanier1974, SamkoKilbasMarichev1974,  MillerRoss1993, Podlubny1998}.
\begin{equation}\label{eq:Eq18}
D^r\1=\frac{t^{-r}}{\Gamma (\1-r)} \comma \enskip r\in \mathbb{R}^+ \enskip \text{and} \enskip t>\0
\end{equation}
For $ r\in \mathbb{N}$, $D^r\1=\0$ due to the poles of the Gamma function at 0, -1, -2 and the classical results are recovered. Clearly, in Eq. \eqref{eq:Eq18} time needs to be positive $(t>\0)$; otherwise, the result of Eq. \eqref{eq:Eq18} would be a complex number when $r\in\left\lbrace\mathbb{R}^+-\mathbb{N}\right\rbrace$. Accordingly, a more formal expression of equation Eq. \eqref{eq:Eq18} within the context of generalized functions is
\begin{equation}\label{eq:Eq19}
D^r U(t-\0)=\frac{\1}{\Gamma(\1-r)}\frac{\1}{t^r}\comma \enskip r\in\mathbb{R}^+ \comma \enskip t>\0
\end{equation}
where $U(t-\0)$ is the Heaviside unit-step function at the time origin \citep{Lighthill1958}.

For the case where $r>\1$, $\1-r=-q$ with $q\in \mathbb{R}^+$; therefore $\1+q=r>\1$. Accordingly, for $r>\1$, Eq. \eqref{eq:Eq19} can be expressed as
\begin{equation}\label{eq:Eq20}
\frac{\mathrm{d}^{\1+q}}{\mathrm{d}t^{\1+q}}U(t-\0) = \frac{\mathrm{d}^q}{\mathrm{d}t^q}\left[ \frac{\mathrm{d}^{\1}U(t-\0)}{\mathrm{d}t} \right] = \frac{\mathrm{d}^q}{\mathrm{d}t^q} \delta(t-\0) =\frac{\1}{\Gamma(-q)}\frac{\1}{t^{\1+q}} \comma \enskip q\in \mathbb{R}^+\comma \enskip t>\0
\end{equation}
and the result of Eq. \eqref{eq:Eq17} is recovered. In Eq. \eqref{eq:Eq20} we used that $\delta(t-\0)=$ {\large$\frac{\mathrm{d}^{\1}U(t-\0)}{\mathrm{d}t}$} \citep{Lighthill1958}.

\section{The Inverse Laplace Transform of $s^q$ with $q \in \mathbb{R}^+$}
\vspace*{-0.5cm}
The memory function, $M(t)$ appearing in Eq. \eqref{eq:Eq07}, of the Scott-Blair $($springpot when $\0 \leq q \leq\1)$ element expressed by Eq. \eqref{eq:Eq09} results directly from the definition of the fractional derivative expressed with the Reimann-Liouville integral given by Eq. \eqref{eq:Eq11}. Substitution of Eq. \eqref{eq:Eq11} into Eq. \eqref{eq:Eq09} gives
\begin{equation}\label{eq:Eq21}
\tau(t)=\frac{\mu_q}{\Gamma(-q)}\int_{\0^-}^t \frac{\gamma(\xi)}{(t-\xi)^{q+\1}} \mathrm{d}\xi \comma \enskip q\in \mathbb{R}^+
\end{equation} 
By comparing Eq. \eqref{eq:Eq21} with Eq. \eqref{eq:Eq07}, the memory function, $M(t)$, of the Scott-Blair element is merely the kernel of the Riemann--Liouville convolution multiplied with the material parameter $\mu_q$
\begin{equation}\label{eq:Eq22}
M(t)=\frac{\mu_q}{\Gamma(-q)}\frac{\1}{t^{q+\1}}=\mu_q\frac{\mathrm{d}^q\delta(t-\0)}{\mathrm{d}t^q}\comma \enskip q\in \mathbb{R}^+
\end{equation}
where the right-hand side of Eq. \eqref{eq:Eq22} is from Eq. \eqref{eq:Eq17}. Equation \eqref{eq:Eq22} shows that the memory function of the springpot element is the fractional derivative of order $q\in \mathbb{R}^+$ of the Dirac delta function as was anticipated by using the argument of physical continuity given that the springpot element interpolates the Hookean spring and the Newtonian dashpot.

In this study we adopt the name ``Scott-Blair element'' rather than the more restrictive ``springpot'' element given that the fractional order of differentiation $q\in \mathbb{R}^+$ is allowed to take values larger than one. The complex dynamic modulus, $\mathcal{G}(\omega)$, of the Scott-Blair fluid described by Eq. \eqref{eq:Eq09} with now $q \in \mathbb{R}^+$ derives directly from Eq. \eqref{eq:Eq12}
\begin{equation}\label{eq:Eq23}
\mathcal{G}(\omega)=\frac{\tau(\omega)}{\gamma(\omega)}=\mu_q(\operatorname{i}\omega)^q
\end{equation}
and its inverse Fourier transform is the memory function, $M(t)$, as indicated by Eq. \eqref{eq:Eq08}. With the introduction of the fractional derivative of the Dirac delta function expressed by Eq. \eqref{eq:Eq16} or \eqref{eq:Eq22}, the definition of the memory function given by Eq. \eqref{eq:Eq08} offers a new (to the best of our knowledge) and useful result regarding the Fourier transform of the function $\mathcal{F}(\omega)=(\operatorname{i}\omega)^q$ with $q \in \mathbb{R}^+$
\begin{equation}\label{eq:Eq24}
\mathcal{F}^{-\1}(\operatorname{i}\omega)^q=\frac{\1}{\2\pi}\int_{-\infty}^{\infty}(\operatorname{i}\omega)^q e^{\operatorname{i}\omega t} \mathrm{d}\omega= \frac{\mathrm{d}^q\delta(t-\0)}{\mathrm{d}t^q}=\frac{\1}{\Gamma(-q)}\frac{\1}{t^{q+\1}} \comma \enskip q \in \mathbb{R}^+ \comma \enskip t>\0
\end{equation}
In terms of the Laplace variable $s=\operatorname{i}\omega$ (see equivalence of Eqs. \eqref{eq:Eq12} and \eqref{eq:Eq13}), Eq. \eqref{eq:Eq24} gives that
\begin{equation}\label{eq:Eq25}
\mathcal{L}^{-\1}\left\lbrace s^q \right\rbrace = \frac{\mathrm{d}^q\delta(t-\0)}{\mathrm{d}t^q}=\frac{\1}{\Gamma(-q)}\frac{\1}{t^{q+\1}}  \comma \enskip q \in \mathbb{R}^+ \comma \enskip t>\0
\end{equation}
where $\mathcal{L}^{-\1}$ indicates the inverse Laplace transform operator \citep{Erdelyi1954, LePage1961, Mainardi2010}. 

When $t>\0$ the right-hand side of Eq. \eqref{eq:Eq24} or \eqref{eq:Eq25} is non-zero only when $q\in\left\lbrace\mathbb{R}^+-\mathbb{N}\right\rbrace$; otherwise it vanishes because of the poles of the Gamma function when $q$ is zero or any positive integer. The validity of Eq. \eqref{eq:Eq24} can be confirmed by investigating its limiting cases. For instance, when $q=\0$, then $(\operatorname{i}\omega)^q=\1$; and Eq. \eqref{eq:Eq24} yields that {\large $\frac{\1}{\2\pi}$}$\displaystyle\int_{-\infty}^{\infty} e^{\operatorname{i}\omega t} \mathrm{d}\omega=\delta(t-\0)$; which is the correct result. When $q=\1$, Eq. \eqref{eq:Eq24} yields that 
{\large $\frac{\1}{\2\pi}$}$\displaystyle\int_{-\infty}^{\infty}\operatorname{i}\omega e^{\operatorname{i}\omega t} \mathrm{d}\omega=$ {\large $\frac{\mathrm{d}\delta(t-\0)}{\mathrm{d}t}$}. Clearly, the function $\mathcal{F}(\omega)=\operatorname{i}\omega$ is not Fourier integrable in the classical sense, yet the result of Eq. \eqref{eq:Eq24} can be confirmed by evaluating the Fourier transform of {\large $\frac{\mathrm{d}\delta(t-\0)}{\mathrm{d}t}$} in association with the properties of the higher-order derivatives of the Dirac delta function given by Eq. \eqref{eq:Eq06}. By virtue of Eq. \eqref{eq:Eq06}, the Fourier transform of {\large $\frac{\mathrm{d}\delta(t-\0)}{\mathrm{d}t}$} is
\begin{equation}\label{eq:Eq26}
\int_{-\infty}^{\infty} \frac{\mathrm{d}\delta(t-\0)}{\mathrm{d}t} e^{-\operatorname{i}\omega t} \mathrm{d}t = -(-\operatorname{i}\omega) e^{-\operatorname{i}\omega \0}=\operatorname{i}\omega
\end{equation}
therefore, the functions $\operatorname{i}\omega$ and {\large $\frac{\mathrm{d}\delta(t-\0)}{\mathrm{d}t}$} are Fourier pairs, as indicated by Eq. \eqref{eq:Eq24}.

More generally, for any $q=n \in \mathbb{N}$, Eq. \eqref{eq:Eq24} yields that {\large $\frac{\1}{\2\pi}$}$\displaystyle\int_{-\infty}^{\infty} (\operatorname{i}\omega)^n e^{\operatorname{i}\omega t} \mathrm{d}\omega=${\large $\frac{\mathrm{d}^n\delta(t-\0)}{\mathrm{d}t^n}$} and by virtue of Eq. \eqref{eq:Eq06}, the Fourier transform of {\large $\frac{\mathrm{d}^n\delta(t-\0)}{\mathrm{d}t^n}$} is 
\begin{equation}\label{eq:Eq27}
\int_{-\infty}^{\infty} \frac{\mathrm{d}^n\delta(t-\0)}{\mathrm{d}t^n} e^{-\operatorname{i}\omega t} \mathrm{d}t=(-\1)^n (-\operatorname{i}\omega)^n = (\operatorname{i}\omega)^n
\end{equation}
showing that the functions $(\operatorname{i}\omega)^n$ and {\large $\frac{\mathrm{d}^n\delta(t-\0)}{\mathrm{d}t^n}$} are Fourier pairs, which is a special result for $q\in \mathbb{N}_{\0}$ of the more general result offered by Eq. \eqref{eq:Eq24}. Consequently, fractional calculus and the memory function of the Scott--Blair element with $q\in\mathbb{R}^+$ offer an alternative avenue to reach the \citet{GelfandShilov1964} definition of the Dirac delta function and its integer-order derivatives given by Eq. \eqref{eq:Eq02}. By establishing the inverse Laplace transform of $s^q$ with $q\in\mathbb{R}^+$ given by Eq. \eqref{eq:Eq25} we proceed by examining the inverse Laplace transform of {\large $\frac{s^q}{(s\mp \lambda)^{\alpha}}$} with $\alpha< q\in\mathbb{R}^+$.

\section{The Inverse Laplace Transform of {\Large $\frac{s^q}{(s\mp \lambda)^{\alpha}}$} with $\alpha< q\in\mathbb{R}^+$}
\vspace*{-0.5cm}
The inverse Laplace transform of $\mathcal{F}(s)=$ {\large $\frac{s^q}{(s\mp \lambda)^{\alpha}}$} with $\alpha< q\in\mathbb{R}^+$ is evaluated with the convolution theorem \citep{LePage1961}
\begin{equation}\label{eq:Eq28}
f(t)=\mathcal{L}^{-\1}\left\lbrace \mathcal{F}(s) \right\rbrace = \mathcal{L}^{-\1}\left\lbrace \mathcal{H}(s)\mathcal{G}(s) \right\rbrace = \int_{\0}^{t}h(t-\xi)g(\xi)\mathrm{d}\xi
\end{equation}
where $h(t)=\mathcal{L}^{-\1}\left\lbrace \mathcal{H}(s) \right\rbrace=\mathcal{L}^{-\1}\left\lbrace s^q \right\rbrace$ given by Eq. \eqref{eq:Eq25} and $g(t)=\mathcal{L}^{-\1}\left\lbrace \mathcal{G}(s)\right\rbrace=\mathcal{L}^{-\1}${\large $\left\lbrace \frac{\1}{(s\mp\lambda)^{\alpha}} \right\rbrace$} $=$ {\large $\frac{\1}{\Gamma(\alpha)}$}$t^{\alpha-\1}e^{\pm\lambda t}$ shown in entry (2) of Table \ref{tab:Table1} \citep{Erdelyi1954} which summarizes selective known inverse Laplace transforms of functions with arbitrary power. Accordingly, Eq. \eqref{eq:Eq28} gives
\begin{table}[t!]
\caption{Known inverse Laplace transforms of irrational functions with an arbitrary power.}
\vspace{6pt}
\setlength{\tabcolsep}{2pt}
\renewcommand{\arraystretch}{2}
\small
\begin{tabularx}{\linewidth}{>{\centering\arraybackslash}m{0.27\linewidth}| >{\centering\arraybackslash}m{0.37\linewidth}| >{\centering\arraybackslash}m{0.33\linewidth}  }	
	\hline \hline 	
	\thead{\white{(1)}\\ \white{(1)}\\ \white{(1)}} & $\mathcal{F}(s)=\mathcal{L}\left\lbrace f(t) \right\rbrace=\displaystyle\int_{\0}^{\infty}f(t)e^{-st} \mathrm{d}t$  &  $f(t)=\mathcal{L}^{-\1}\left\lbrace \mathcal{F}(s) \right\rbrace$ \tabularnewline 
		\hline
		
	\thead{\white{(1)}\\(1)\\ \white{(1)}} & {\large $\frac{\1}{s^{\alpha}}$} $\quad \alpha\in\mathbb{R}^+$ & {\large $\frac{t^{\alpha-\1}}{\Gamma(\alpha)}$} \tabularnewline \hline 
	
	\thead{\white{(2)}\\(2)\\ \white{(2)}} & {\large $\frac{\1}{(s \mp \lambda)^{\alpha}}$} $\quad \alpha\in\mathbb{R}^+$ & {\large $\frac{\1}{\Gamma(\alpha)}$}$t^{\alpha-\1}e^{\pm\lambda t}$ \tabularnewline \hline
	
	\thead{\white{(3)}\\(3)\\ \white{(3)}} & {\large $\frac{\1}{s^{\alpha}(s \mp \lambda)}$}$\quad \alpha\in\mathbb{R}^+$ & $t^{\alpha}E_{\1\comma\, \1+\alpha}(\pm \lambda t)=I^{\alpha}e^{\pm \lambda t}$ \tabularnewline \hline
	
	\thead{\white{(4)}\\(4)\\ \white{(4)}} & {\large $\frac{s^{\alpha}}{s \mp \lambda}$}$\quad \0<\alpha<\1$  & $t^{-\alpha} E_{\1\comma\, \1-\alpha}(\pm\lambda t)=$ {\large $\frac{\mathrm{d}^{\alpha} e^{\pm \lambda t}}{\mathrm{d}t^{\alpha}}$} \tabularnewline \hline
	
	\thead{\white{(5)}\\(5)\\ \white{(5)}} & {\large $\frac{s^{\alpha-\beta}}{s^{\alpha} \mp \lambda}$}$\quad \alpha\comma\,\beta\in\mathbb{R}^+$  & $t^{\beta-\1} E_{\alpha\comma\, \beta}(\pm\lambda t^{\alpha})$ \tabularnewline \hline
	
\thead{(6)\\ Special case of (5)\\ for $\beta=\1$}	& {\large $\frac{s^{\alpha-\1}}{s^{\alpha} \mp \lambda}$}$\quad \alpha\in\mathbb{R}^+$ & $ E_{\alpha}(\pm\lambda t^{\alpha})$ \tabularnewline \hline

\thead{(7)\\ Special case of (5)\\for $\alpha=\beta$}	& {\large $\frac{\1}{s^{\alpha} \mp \lambda}$}$\quad \alpha\in\mathbb{R}^+$ & $t^{\alpha-\1} E_{\alpha\comma\, \alpha}(\pm\lambda t^{\alpha})=$ {\large $\varepsilon$}$_{\alpha-\1}(\pm\lambda\comma\, t)$ \tabularnewline \hline

\thead{(8)\\ Special case of (5)\\ $\alpha-\beta=-\1$}	& {\large $\frac{\1}{s(s^{\alpha} \mp \lambda)}$}$\quad \alpha\in\mathbb{R}^+$ & $t^{\alpha} E_{\alpha\comma\, \alpha+\1}(\pm\lambda t^{\alpha})$ \tabularnewline \hline

\thead{(9)\\ Special case of (5)\\with $\0<\alpha-\beta=q<\alpha$}	& {\large $\frac{s^q}{s^{\alpha} \mp \lambda}$}$\quad \0<q<\alpha\in\mathbb{R}^+$ & $t^{\alpha-q-\1} E_{\alpha\comma\, \alpha-q}(\pm\lambda t^{\alpha})$ \tabularnewline 
	
		\hline \hline 	
	\end{tabularx}
\label{tab:Table1}
\end{table}
\begin{equation}\label{eq:Eq29}
\mathcal{L}^{-\1} \left\lbrace \frac{s^q}{(s\mp\lambda)^{\alpha}} \right\rbrace = \frac{\1}{\Gamma(-q)} \int_{\0}^{t} \frac{\1}{(t-\xi)^{q+\1}} \frac{\1}{\Gamma(\alpha)}\xi^{\alpha-\1}e^{\pm\lambda \xi}\mathrm{d}\xi
\end{equation}
With reference to Eq. \eqref{eq:Eq11}, Eq. \eqref{eq:Eq29} is expressed as
\begin{equation}\label{eq:Eq30}
\mathcal{L}^{-\1} \left\lbrace \frac{s^q}{(s\mp\lambda)^{\alpha}} \right\rbrace = \frac{\1}{\Gamma(\alpha)} \frac{\mathrm{d}^q}{\mathrm{d}t^q}\left[ t^{\alpha-\1}e^{\pm\lambda t} \right] \comma \enskip \alpha\comma\, q\in\mathbb{R}^+
\end{equation}
For the special case where $\lambda=\0$ and after using that {\large $\frac{\mathrm{d}^q t^{\alpha-\1}}{\mathrm{d}t^q}$} $=$ {\large $\frac{\Gamma(\alpha)}{\Gamma(\alpha-q)}$}$t^{\alpha-\1-q}$ \citep{MillerRoss1993}, Eq. \eqref{eq:Eq30} reduces to
\begin{equation}\label{eq:Eq31}
\mathcal{L}^{-\1} \left\lbrace s^{q-\alpha} \right\rbrace = \frac{\1}{\Gamma(\alpha)}\frac{\mathrm{d}^q}{\mathrm{d}t^q}t^{\alpha-\1}=\frac{\1}{\Gamma(\alpha)}\frac{\Gamma(\alpha)}{\Gamma(-q+\alpha)}\frac{\1}{t^{q-\alpha+\1}}=\frac{\mathrm{d}^{q-\alpha}\delta(t-\0)}{\mathrm{d}t^{q-\alpha}}\comma \enskip \alpha<q\in\mathbb{R}^+
\end{equation}
and the result of Eq. \eqref{eq:Eq25} is recovered. Equation \eqref{eq:Eq31} also reveals the intimate relation between the fractional derivative of the Dirac delta function and the fractional derivative of the power law
\begin{equation}\label{eq:Eq32}
\frac{\mathrm{d}^{q-\alpha}\delta(t-\0)}{\mathrm{d}t^{q-\alpha}}=\frac{\1}{\Gamma(\alpha)}\frac{\mathrm{d}^q}{\mathrm{d}t^q}t^{\alpha-\1} \comma \enskip \alpha< q\in\mathbb{R}^+
\end{equation}
For the special case where $q=\alpha$, Eq. \eqref{eq:Eq32} yields $\delta(t-\0)=$ {\large $\frac{\1}{\Gamma(\alpha)}\frac{\Gamma(\alpha)}{\Gamma(\0)}$}$t^{-\1}=$ {\large $\frac{\1}{\Gamma(0)}\frac{\1}{t}$} and the \citet{GelfandShilov1964} definition of the Dirac delta function given by Eq. \eqref{eq:Eq02} is recovered. The new results, derived in this paper, on the inverse Laplace transform of irrational functions with arbitrary powers are summarized in Table \ref{tab:Table2}.

\section{The Inverse Laplace Transform of {\Large $\frac{s^q}{s^{\alpha}\mp \lambda}$} with $\alpha\comma\, q\in\mathbb{R}^+$}
\vspace*{-0.5cm}
We start with the known result for the inverse Laplace transform of the function $\mathcal{Q}(s)=$ {\large $\frac{s^{\alpha-\beta}}{s^{\alpha}\mp \lambda}$} with $\alpha\comma\, \beta \in \mathbb{R^+}$ \citep{GorenfloMainardi1997, Podlubny1998}
\begin{equation}\label{eq:Eq33}
\mathcal{L}^{-\1}\left\lbrace \frac{s^{\alpha-\beta}}{s^{\alpha}\mp \lambda} \right\rbrace = t^{\beta-1} E_{\alpha\comma\, \beta}(\pm \lambda t^{\alpha}) \comma \enskip \lambda\comma\, \alpha\comma\, \beta \in \mathbb{R}^+
\end{equation}
where $E_{\alpha\comma\, \beta}(z)$ is the two-parameter Mittag--Leffler function \citep{Erdelyi1953, HauboldMathaiSaxena2011, GorenfloKilbasMainardiRogosin2014}
\begin{equation}\label{eq:Eq34}
E_{\alpha\comma\, \beta}(z)= \sum_{j=0}^{\infty}\frac{z^j}{\Gamma(j\alpha+\beta)} \comma\, \enskip \alpha\comma\, \beta > \0
\end{equation}
When $\beta=\1$, Eq. \eqref{eq:Eq33} reduces to the result of the Laplace transform of the one-parameter Mittag--Leffler function, originally derived by Mittag--Leffler \citep{GorenfloKilbasMainardiRogosin2014}
\begin{equation}\label{eq:Eq35}
\mathcal{L}^{-\1}\left\lbrace \frac{s^{\alpha-\1}}{s^{\alpha}\mp \lambda} \right\rbrace = E_{\alpha\comma\,\1}(\pm \lambda t^{\alpha}) =  E_{\alpha}(\pm \lambda t^{\alpha}) \comma \enskip \lambda\comma\,\alpha\in\mathbb{R}^+
\end{equation}
\begin{figure}[b!]
\centering
\includegraphics[width=\linewidth, angle=0]{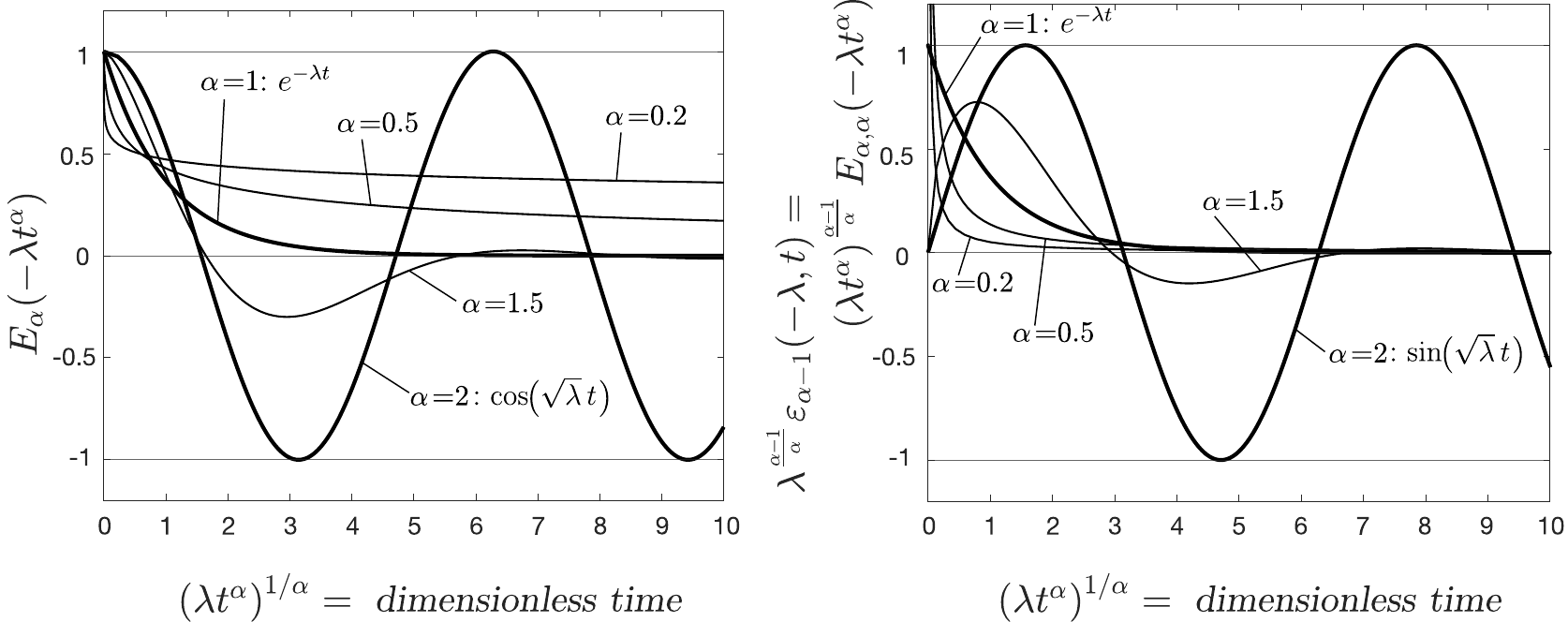}
\caption{The one-parameter Mittag--Leffler function $E_{\alpha}(-\lambda t^{\alpha})$ (left) and the Rabotnov function {\large $\varepsilon$}$_{\alpha-\1}(-\lambda\comma\, t)=t^{\alpha-\1}E_{\alpha\comma\,\alpha}(-\lambda t^{\alpha})$ (right) for various values of the parameter $\alpha\in\mathbb{R}^+$.}
\label{fig:Fig02}
\end{figure}
When $\alpha=\beta$, the right-hand side of Eq. \eqref{eq:Eq33} is known as the Rabotnov function, {\large $\varepsilon$}$_{\alpha-\1}(\pm \lambda\comma\, t)=t^{\alpha-\1} E_{\alpha\comma\,\alpha}(\pm \lambda t^{\alpha})$ \citep{Rabotnov1980, Mainardi2010, Makris2020, MakrisEfthymiou2020}; and Eq. \eqref{eq:Eq33} yields
\begin{equation}\label{eq:Eq36}
\mathcal{L}^{-\1}\left\lbrace \frac{\1}{s^{\alpha}\mp \lambda} \right\rbrace = t^{\alpha-\1} E_{\alpha\comma\,\alpha}(\pm \lambda t^{\alpha})=\text{{\large $\varepsilon$}}_{\alpha-\1}(\pm \lambda\comma\, t) \comma \enskip \lambda\comma\,\alpha\in\mathbb{R}^+
\end{equation}
Figure \ref{fig:Fig02} plots the function $E_{\alpha}(- \lambda t^{\alpha})$ (left) and the function \text{{\large $\varepsilon$}}$_{\alpha-\1}(- \lambda\comma\, t)=t^{\alpha-\1} E_{\alpha\comma\,\alpha}(- \lambda t^{\alpha})$ (right) for various values of the parameter $\alpha\in\mathbb{R}^+$. For $\alpha=\1$ both functions contract to $e^{-\lambda t}$. When $\alpha-\beta=-\1$ Eq. \eqref{eq:Eq33} gives:
\begin{equation}\label{eq:Eq37}
\mathcal{L}^{-\1}\left\lbrace \frac{\1}{s(s^{\alpha}\mp \lambda)} \right\rbrace = t^{\alpha} E_{\alpha\comma\,\alpha+\1}(\pm\lambda t^{\alpha})  \comma \enskip \lambda\comma\,\alpha\in\mathbb{R}^+
\end{equation}

\begin{table}[t!]
\caption{New results on the inverse Laplace transform of irrational functions with arbitrary powers.}
\vspace{6pt}
\setlength{\tabcolsep}{2pt}
\small 
\singlespacing
\begin{tabularx}{\linewidth}{>{\centering\arraybackslash}m{0.20\linewidth}| >{\centering\arraybackslash}m{0.35\linewidth}| >{\centering\arraybackslash}m{0.42\linewidth}  }	
	\hline \hline 	
	\thead{\white{(1)}\\ \white{(1)}\\ \white{(1)}}  &  $\mathcal{F}(s)=\mathcal{L}\left\lbrace f(t) \right\rbrace=\displaystyle\int_{\0}^{\infty}f(t)e^{-st} \mathrm{d}t$  & $f(t)=\mathcal{L}^{-\1}\left\lbrace \mathcal{F}(s) \right\rbrace$ \tabularnewline 
		\hline
		
	\thead{\white{(1)}\\ (1)\\ \white{(1)}} &$s^q \quad q\in\mathbb{R}^+$ & {\large $\frac{\1}{\Gamma(-q)}\frac{\1}{t^{q+\1}}$} $=$ {\large $\frac{\mathrm{d}^q\delta(t-\0)}{\mathrm{d}t^q}$} \tabularnewline \hline
	
	\thead{\white{(2)}\\ (2)\\ \white{(2)}} & {\large $\frac{s^q}{(s\mp \lambda)^{\alpha}}$} $\quad \alpha\comma\,q\in\mathbb{R}^+$ & {\large $\frac{\1}{\Gamma(\alpha)} \, \frac{\mathrm{d}^q}{\mathrm{d}t^q}$}$\left[ t^{\alpha-\1}e^{\pm \lambda t} \right]$  \tabularnewline \hline
	
	\thead{(3)\\ Extension of entry\\(9) of Table \ref{tab:Table1} for\\$\alpha<q<\2\alpha\in\mathbb{R}^+$} &  {\large $\frac{s^q}{s^{\alpha}\mp \lambda}$}$\quad \alpha<q<\2\alpha\in\mathbb{R}^+$ & \thead{{\large $\frac{\1}{\Gamma(-q+\alpha)}\frac{\1}{t^{q-\alpha+\1}}$}$\pm \lambda t^{\2\alpha-q-\1}E_{\alpha\comma\, \2\alpha-q}(\pm \lambda t^{\alpha})$\\ \\$=${\large $\frac{\mathrm{d}^{q-\alpha}}{\mathrm{d}t^{q-\alpha}}$}$\delta(t-\0) \pm \lambda t^{\2\alpha-q-\1}E_{\alpha\comma\, \2\alpha-q}(\pm \lambda t^{\alpha})$} \tabularnewline \hline
	
	\thead{(4)\\ Special case of (3)\\ for $\alpha=\1$} &  {\large $\frac{s^q}{s \mp \lambda}$}$\quad \1<q<\2$ & \thead{{\large $\frac{\1}{\Gamma(-q+\1)}\frac{\1}{t^q}$}$\pm \lambda t^{\1-q}E_{\1\comma\, \2-q}(\pm \lambda t)\quad \quad \quad \quad \quad$\\ \\$\quad \quad \quad \quad =$ {\large $\frac{\mathrm{d}^{q-\1}\delta(t-\0)}{\mathrm{d}t^{q-\1}}$}$\pm\lambda t^{\1-q}E_{\1\comma\, \2-q}(\pm\lambda t)$} \tabularnewline \hline	
			
	\thead{(5)\\General case of (3)\\for any $q\in\mathbb{R}^+$ with\\ $n\alpha<q<(n+\1)\alpha\comma\,$\\$ n\in\mathbb{N}$} &  {\large $\frac{s^q}{s^{\alpha}\mp \lambda}$}$\quad \alpha<q\in\mathbb{R}^+$ & {\footnotesize \thead{$\displaystyle \sum_{j=\1}^{n}(\pm\lambda)^{j-\1}${\large $\frac{\1}{\Gamma(-q+j\alpha)}\frac{\1}{t^{q-j\alpha+\1}}$}$+ \quad \quad \quad \quad $\\$ \quad \quad (\pm \lambda)^nt^{(n+\1)\alpha-q-\1}E_{\alpha\comma\, (n+\1)\alpha-q}(\pm \lambda t^{\alpha})$\\ \\$=\displaystyle \sum_{j=\1}^{n}(\pm\lambda)^{j-\1}${\large $\frac{\mathrm{d}^{q-j\alpha}}{\mathrm{d}t^{q-j\alpha}}$}$\delta(t-\0)+ \quad \quad \quad \quad \quad \quad$\\$\quad \quad (\pm \lambda)^nt^{(n+\1)\alpha-q-\1}E_{\alpha\comma\, (n+\1)\alpha-q}(\pm \lambda t^{\alpha})\comma$\\ $\quad \quad \quad \quad \quad \quad \quad \quad \quad \quad n\alpha<q<(n+\1)\alpha$\\$ =$ {\large $\frac{\mathrm{d}^q}{\mathrm{d}t^q}\varepsilon$}$_{\alpha-\1}(\pm\lambda\comma\, t)\quad \quad \quad \quad \quad \quad \quad \quad \quad \quad \quad $ }} \tabularnewline \hline

	\thead{(6)\\Special case of (5)\\for $\alpha=\1$ with\\ $n<q<n+\1\comma$\\$ n\in\mathbb{N}$} &  {\large $\frac{s^q}{s\mp \lambda}$}$\quad \1<q\in\mathbb{R}^+$ & {\footnotesize \thead{$\displaystyle \sum_{j=\1}^{n}(\pm\lambda)^{j-\1}${\large $\frac{\1}{\Gamma(-q+j)}\frac{\1}{t^{q-j+\1}}$}$+ \quad \quad \quad \quad $\\$ \quad \quad \quad \quad \quad \quad \quad \quad (\pm \lambda)^nt^{n-q}E_{\1\comma\, n+\1-q}(\pm \lambda t)$\\ \\$=\displaystyle \sum_{j=\1}^{n}(\pm\lambda)^{j-\1}${\large $\frac{\mathrm{d}^{q-j}}{\mathrm{d}t^{q-j}}$}$\delta(t-\0)+ \quad \quad \quad \quad \quad$\\$\quad \quad \quad \quad \quad \quad \quad \quad (\pm \lambda)^nt^{n-q}E_{\1\comma\, n+\1-q}(\pm \lambda t)\comma$\\$\quad \quad \quad \quad \quad \quad \quad \quad \quad \quad \quad \quad \quad n<q<n+\1$\\ $ =$ {\large $\frac{\mathrm{d}^q}{\mathrm{d}t^q}$}$e^{\pm\lambda t} \quad \quad \quad \quad \quad \quad \quad \quad \quad \quad \quad \quad \quad \quad $ }} \tabularnewline
	
		\hline \hline 	
	\end{tabularx}
\label{tab:Table2}
\end{table}

The inverse Laplace transform of $\mathcal{F}(s)=$ {\large $\frac{s^q}{s^{\alpha} \mp \lambda}$} with $\alpha\comma\, q \in \mathbb{R}^+$ is evaluated with the convolution theorem expressed by Eq. \eqref{eq:Eq28}
where $h(t)=\mathcal{L}^{-\1}\left\lbrace \mathcal{H}(s) \right\rbrace=\mathcal{L}^{-\1}\left\lbrace s^q \right\rbrace$ given by Eq. \eqref{eq:Eq25} and $g(t)=$ {\large $\varepsilon$}$_{\alpha-\1}(\pm \lambda\comma\, t)$ is given by Eq. \eqref{eq:Eq36}. Accordingly, Eq. \eqref{eq:Eq28} gives
\begin{equation}\label{eq:Eq38}
\mathcal{L}^{-\1}\left\lbrace \frac{s^q}{s^{\alpha}\mp \lambda} \right\rbrace = \frac{\1}{\Gamma(-q)}\int_{\0}^{t}\frac{\1}{(t-\xi)^{q+\1}}\xi^{\alpha-\1}E_{\alpha\comma\,\alpha}(\pm \lambda\xi^{\alpha})\mathrm{d}\xi
\end{equation}
With reference to Eq. \eqref{eq:Eq11}, Eq. \eqref{eq:Eq38} indicates that {\large {\normalsize $\mathcal{L}^{-\1}$}$\left\lbrace \frac{s^q}{s^{\alpha}\mp \lambda} \right\rbrace$} is the fractional derivative of order $q$ of the Rabotnov function {\large $\varepsilon$}$_{\alpha-\1}(\pm \lambda\comma\, t)$
\begin{equation}\label{eq:Eq39}
\mathcal{L}^{-\1}\left\lbrace \frac{s^q}{s^{\alpha}\mp \lambda} \right\rbrace = \frac{\mathrm{d}^q}{\mathrm{d}t^q}\left[ t^{\alpha-\1} E_{\alpha\comma\,\alpha}(\pm \lambda t^{\alpha}) \right] = t^{\alpha-q-\1} E_{\alpha\comma\,\alpha-q}(\pm \lambda t^{\alpha}) 
\end{equation}
For the case where $q<\alpha\in \mathbb{R}^+$, the exponent $q$ can be expressed as $q=\alpha-\beta$ with $\0<\beta\leq\alpha\in\mathbb{R}^+$ and Eq. \eqref{eq:Eq39} returns the known result given by Eq. \eqref{eq:Eq33}. For the case where $q>\alpha\in\mathbb{R}^+$, the numerator of the fraction {\large $\frac{s^q}{s^{\alpha}\mp \lambda}$} is more powerful than the denominator and the inverse Laplace transform expressed by Eq. \eqref{eq:Eq39} is expected to yield a singularity which is manifested with the second parameter of the Mittag--Leffler function $E_{\alpha\comma\,\alpha-q}(\pm\lambda t^{\alpha})$, being negative $(\alpha-q<\0)$. This embedded singularity in the right-hand side of Eq. \eqref{eq:Eq39} when $q>\alpha$ is extracted by using the recurrence relation \citep{Erdelyi1953, HauboldMathaiSaxena2011, GorenfloKilbasMainardiRogosin2014}
\begin{equation}\label{eq:Eq40}
E_{\alpha\comma\,\beta}(z)=\frac{\1}{\Gamma(\beta)}+zE_{\alpha\comma\,\alpha+\beta}(z)
\end{equation}
By employing the recurrence relation \eqref{eq:Eq40} to the right-hand side of Eq. \eqref{eq:Eq39}, then Eq. \eqref{eq:Eq39}  for $q>\alpha\in\mathbb{R}^+$ assumes the expression
\begin{equation}\label{eq:Eq41}
\mathcal{L}^{-\1}\left\lbrace \frac{s^q}{s^{\alpha}\mp \lambda} \right\rbrace = \frac{\1}{\Gamma(-q+\alpha)}\frac{\1}{t^{q-\alpha+\1}}\pm\lambda t^{2\alpha-q-\1} E_{\alpha\comma\, 2\alpha-q}(\pm \lambda t^{\alpha})
\end{equation}
Recognizing that according to Eq. \eqref{eq:Eq17}, the first term in the right-hand side of Eq. \eqref{eq:Eq40} is {\large $\frac{\mathrm{d}^{q-\alpha}\delta(t-\0)}{\mathrm{d}t^{q-\alpha}}$}, the inverse Laplace transform of {\large $\frac{s^q}{s^{\alpha}\mp \lambda} $} with $q>\alpha\in\mathbb{R}^+$ can be expressed in the alternative form
\begin{equation}\label{eq:Eq42}
\mathcal{L}^{-\1}\left\lbrace \frac{s^q}{s^{\alpha}\mp \lambda} \right\rbrace = \frac{\mathrm{d}^{q-\alpha}}{\mathrm{d}t^{q-\alpha}} \delta(t-\0) \pm \lambda t^{\2\alpha-q-\1} E_{\alpha\comma\, \2\alpha-q}(\pm \lambda t^{\alpha}) \comma \enskip \alpha<q<\2\alpha
\end{equation}
in which the singularity {\large $\frac{\mathrm{d}^{q-\alpha}\delta(t-\0)}{\mathrm{d}t^{q-\alpha}}$} has been extracted from the right-hand side of Eq. \eqref{eq:Eq39}, and now the second index of the Mittag--Leffler function appearing in Eq. \eqref{eq:Eq41} or \eqref{eq:Eq42} has been increased to $\2\alpha-q$. In the event that $\2\alpha-q$ remains negative $(q>\2\alpha)$, the Mittag--Leffler function appearing on the right-hand side of Eq. \eqref{eq:Eq41} or \eqref{eq:Eq42} is replaced again by virtue of the recurrence relation  \eqref{eq:Eq40} and results in
\begin{equation}\label{eq:Eq43}
\mathcal{L}^{-\1}\left\lbrace \frac{s^q}{s^{\alpha}\mp \lambda} \right\rbrace = \frac{\mathrm{d}^{q-\alpha}}{\mathrm{d}t^{q-\alpha}} \delta(t-\0) \pm \lambda \frac{\mathrm{d}^{q-\2\alpha}}{\mathrm{d}t^{q-\2\alpha}} \delta(t-\0) + (\pm \lambda)^{\2} t^{\3\alpha-q-\1} E_{\alpha\comma\, \3\alpha-q}(\pm \lambda t^{\alpha}) \comma \enskip \2\alpha<q<\3\alpha
\end{equation}
More generally, for any $q\in\mathbb{R}^+$ with $n\alpha<q<(n+\1)\alpha$ with $n\in\mathbb{N}=\left\lbrace \1\comma\, \2\comma\, ... \right\rbrace$ and $\alpha\in\mathbb{R}^+$
\begin{align}\label{eq:Eq44}
\mathcal{L}^{-\1}\left\lbrace \frac{s^q}{s^{\alpha}\mp \lambda} \right\rbrace = & \frac{\mathrm{d}^q}{\mathrm{d}t^q}\text{{\large $\varepsilon$}}_{\alpha-\1}(\pm\lambda\comma\,t) = \\ \nonumber
& \sum_{j=\1}^{n} (\pm \lambda)^{j-\1}\frac{\mathrm{d}^{q-j\alpha}}{\mathrm{d}t^{q-j\alpha}} \delta(t-\0)  + (\pm \lambda)^n t^{(n+\1)\alpha-q-\1} E_{\alpha\comma\, (n+\1)\alpha-q}(\pm \lambda t^{\alpha}) 
\end{align}
and all singularities from the Mittag--Leffler function have been extracted. For the special case where $\alpha=\1$ Eq. \eqref{eq:Eq44} gives for $n<q<n+\1$ with $n\in\mathbb{N}=\left\lbrace \1\comma\, \2\comma\, ... \right\rbrace$
\begin{figure}[b!]
\centering
\includegraphics[width=0.8\linewidth, angle=0]{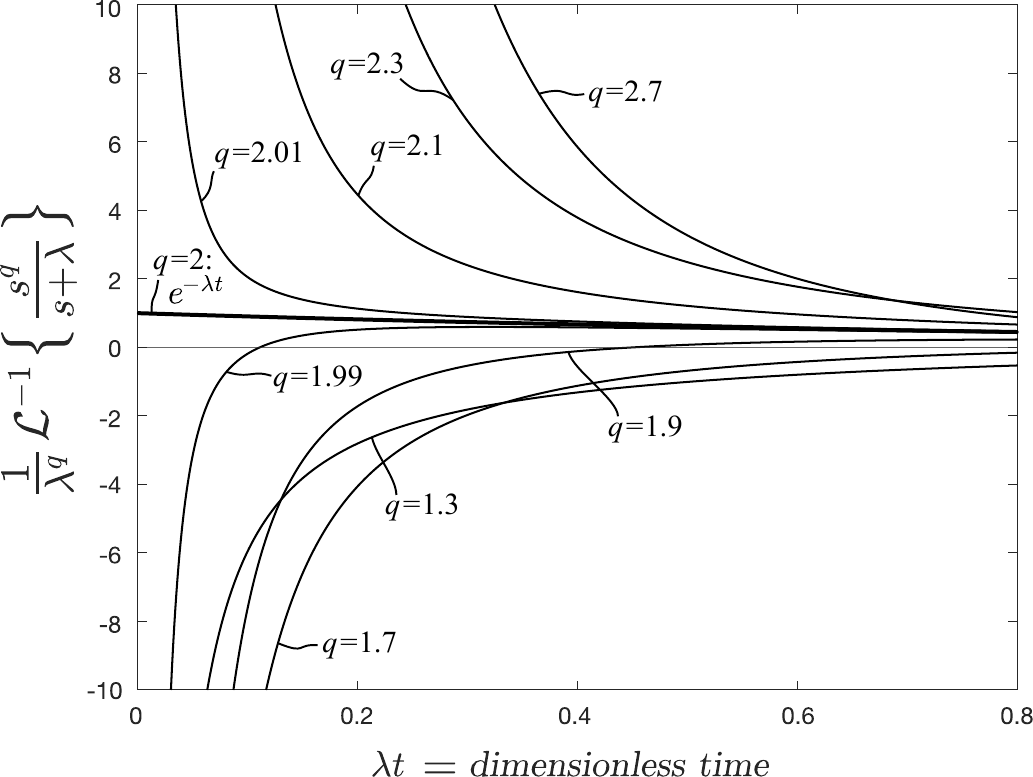}
\caption{Plots of $\mathcal{L}^{-\1}${\large $\left\lbrace \frac{s^q}{s+ \1} \right\rbrace$} $=$ {\large $\frac{\mathrm{d}^q}{\mathrm{d}t^q}$}$e^{-t}$ by using Eq. \eqref{eq:Eq46} for $\1<q<\2$ and Eq. \eqref{eq:Eq47} for $\2<q<\3$. When $q$ tends to 2 from below, the curves for {\large $\frac{\1}{\lambda^q}\mathcal{L}^{-\1}\left\lbrace \frac{s^q}{s + \lambda} \right\rbrace$} approach $e^{-\lambda t}$ from below; whereas when $q$ tends to 2 from above, the curves of the inverse Laplace transform approach $e^{-\lambda t}$ from above.}
\label{fig:Fig03}
\end{figure}
\begin{equation}\label{eq:Eq45}
\mathcal{L}^{-\1}\left\lbrace \frac{s^q}{s\mp \lambda} \right\rbrace = \frac{\mathrm{d}^q}{\mathrm{d}t^q}e^{\pm\lambda t} = \sum_{j=\1}^{n} (\pm \lambda)^{j-\1}\frac{\mathrm{d}^{q-j}}{\mathrm{d}t^{q-j}} \delta(t-\0)  + (\pm \lambda)^n t^{n-q} E_{\1\comma\, n+\1-q}(\pm \lambda t) 
\end{equation}
which is the extension of entry (4) of Table \ref{tab:Table1} for any $q\in\mathbb{R}^+$. As an example, for $\1<q<\2$ Eq. \eqref{eq:Eq45} is expressed in its dimensionless form
\begin{equation}\label{eq:Eq46}
\frac{\1}{\lambda^q}\mathcal{L}^{-\1}\left\lbrace \frac{s^q}{s + \lambda} \right\rbrace = \frac{\1}{\Gamma(-q+\1)}\frac{\1}{(\lambda t)^q}-\frac{\1}{(\lambda t)^{q-\1}}E_{\1\comma\,\2-q}(-\lambda t)\comma \enskip \1<q<\2
\end{equation}
whereas for $\2<q<\3$, Eq. \eqref{eq:Eq45} yields
\begin{equation}\label{eq:Eq47}
\frac{\1}{\lambda^q}\mathcal{L}^{-\1}\left\lbrace \frac{s^q}{s + \lambda} \right\rbrace = \frac{\1}{\Gamma(-q+\1)}\frac{\1}{(\lambda t)^q}-\frac{\1}{\Gamma(-q+\2)}\frac{\1}{(\lambda t)^{q-\1}}+\frac{\1}{(\lambda t)^{q-\2}}E_{\1\comma\,\3-q}(-\lambda t)\comma \enskip \2<q<\3
\end{equation}
Figure \ref{fig:Fig03} plots the results of Eq. \eqref{eq:Eq46} for $q=$ 1.3, 1.7, 1.9 and 1.99 together with the results of Eq. \eqref{eq:Eq47} for $q=$ 2.01, 2.1, 2.3 and 2.7. When $q$ tends to 2 from below, the curves for {\large $\frac{\1}{\lambda^q}\mathcal{L}^{-\1}\left\lbrace \frac{s^q}{s + \lambda} \right\rbrace$} approach $e^{-\lambda t}$ from below; whereas when $q$ tends to 2 from above, the curves of the inverse Laplace transform approach $e^{-\lambda t}$ from above.

\section{Summary}
\vspace*{-0.5cm}
In this paper we first show that the memory function, $M(t)$, of the fractional Scott--Blair fluid, $\tau(t)=\mu_q${\large $\frac{\mathrm{d}^q \gamma(t)}{\mathrm{d}t^q}$} with $q\in\mathbb{R}^+$ $($springpot when $\0\leq q\leq\1)$ is the fractional derivative of the Dirac delta function {\large $\frac{\mathrm{d}^q \delta(t-\0)}{\mathrm{d}t^q}$} with $q\in\mathbb{R}^+$. Given that the memory function $M(t)=$ {\large $\frac{\1}{\2\pi}$}$\displaystyle\int_{-\infty}^{\infty}\mathcal{G}(\omega)e^{\operatorname{i}\omega t}\mathrm{d}t$ is the inverse Fourier transform of the complex dynamic modulus, $\mathcal{G}(\omega)$, in association with that $M(t)$ is causal $(M(t)=\0$ for $t<\0)$ we showed that the inverse Laplace transform of $s^q$ for any $q\in\mathbb{R}^+$ is the fractional derivative of order $q$ of the Dirac delta function. This new finding in association with the convolution theorem makes possible the calculation of the inverse Laplace transform of {\large $\frac{s^q}{s^{\alpha}\mp \lambda}$} when $\alpha<q\in\mathbb{R}^+$ which is the fractional derivative of order $q$ of the Rabotnov function {\large $\varepsilon$}$_{\alpha-\1}(\pm\lambda\comma\, t)=t^{\alpha-\1}E_{\alpha\comma\,\alpha}(\pm\lambda t^{\alpha})$. The fractional derivative of order $q\in\mathbb{R}^+$ of the Rabotnov function {\large $\varepsilon$}$_{\alpha-\1}(\pm\lambda\comma\, t)$ produces singularities which are extracted with a finite number of fractional derivatives of the Dirac delta function depending on the strength of the order of differentiation $q$ in association with the recurrence formula of the two-parameter Mittag--Leffler function.

%\clearpage
%
%: $e^{-\lambda t}$
%
%: $\cos(\sqrt{\lambda}\,t)$
%
%: $\sin(\sqrt{\lambda}\,t)$
%
%\clearpage

\bibliographystyle{myapalike}
\bibliography{References} % Path to your References.bib file
\vspace*{-0.5cm}
\end{document}